\documentclass[11pt,twoside]{article}
\usepackage{asp2004}
\usepackage{psfig}
\usepackage{epsf}
\usepackage{graphics}
\usepackage{lscape}
\def\edcomment#1{\iffalse\marginpar{\raggedright\sl#1\/}\else\relax\fi}
\reversemarginpar
\def\msun{\mbox{$M_{\odot}$}}
\def\reff{\mbox{$R_{\rm eff}$}}

\begin{document}
\title{Young Massive Star Clusters in Normal Galaxies}
\author{S. S. Larsen}
\affil{ESO / ST-ECF, Karl-Schwarzschild-Str.\ 2, D-85748 Garching bei
 M{\"u}nchen, Germany}

\begin{abstract}
  Young star clusters with masses similar to those of classical old globular 
clusters are observed not only in starbursts, mergers or otherwise disturbed 
galaxies, but also in normal spiral galaxies. Some young clusters with masses 
as high as $\sim10^6 \, \msun$ have been found in the disks of isolated 
spirals. Dynamical mass estimates are available for a few of these
clusters and are consistent with Kroupa-type IMFs.  The luminosity (and 
possibly mass-) functions of young clusters are usually well approximated by 
power-laws. Thus, massive clusters at the tail of the distribution are 
naturally rare, but appear to be present whenever clusters form in large 
numbers.  While bound star clusters may generally form with a higher 
efficiency in environments of high star formation rate, many of the apparent 
differences between clusters in starbursts and ``normal'' galaxies might 
be simply due to sampling effects. 
\end{abstract}
\thispagestyle{plain}


  It is a human habit to characterise those things with which we are most 
familiar as normal. Although large spiral galaxies are not the 
most common type of galaxy in the Universe, we happen to live within one 
and many astronomers would probably tend to characterise the Milky Way 
as a fairly normal galaxy. Thus, at least for the purpose of this paper, 
``normal'' galaxies mostly refer to non-interacting star forming disk 
galaxies.  Our location within the Milky Way gives us a unique perspective 
from which we can study many of its properties in great detail, and it 
naturally provides a benchmark for comparison with other galaxies.  
Nevertheless, we should question whether it is justified to apply 
results obtained from studies of our own Galaxy to other galaxies which may 
appear superficially similar to it.  In the context of this workshop, it is 
of particular relevance to ask how similar the cluster system in the Milky Way 
is to those in other galaxies. An increasing amount of observational
evidence is pointing to the conclusion that many spirals host
``young massive clusters'' (YMCs) or ``super star clusters'' similar to those
observed in large numbers in starburst galaxies. The definition of a YMC is 
rather vague and varies from one author to another, but the term generally
seems to refer to young clusters that are more massive than the most massive
open clusters in the Milky Way.  However, giving a meaningful definition of 
a massive cluster may eventually be as difficult as distinguishing galaxies 
that are normal from those that are not. 

\section{A bit of background}

  It may be worth recalling some of the main properties of the Milky Way 
open cluster system. The census of open clusters is still highly incomplete 
beyond distances of a few kpc from the Sun, although the situation is 
improving with new surveys such as 2MASS (see e.g.\ the contributions by 
Carpenter and Hanson in this volume). The luminosity function of Milky Way 
open clusters was analysed by \citet{van84}, who found it to be well modelled 
by a power-law $N(L)dL \propto L^{-1.5}dL$ over the range $-8 < M_V < -3$.  
However, they also noted that 
extrapolation of this luminosity function would predict about 100 clusters as 
bright as $M_V=-11$ in the Galaxy, clearly at odds with observations, and thus 
suggested some flattening of the LF slope at higher luminosities.  The 
brightest known young clusters (e.g.\ NGC~3603, $h$ and $\chi$ Per) have 
absolute $V$ magnitudes of $M_V\sim-10$, corresponding to total masses of 
several thousand \msun . Recently, there have been claims that the
Cyg OB2 association might be an even more massive cluster \citep{kno00},
but this object is probably too diffuse to be a bound star cluster 
(though it does have 
a compact core). There are, however, a number of old ($>1$ Gyr) open clusters 
in the Milky Way with masses of $\sim10^4$ \msun\
\citep{friel95}. These objects are likely to have lost a significant fraction
of their total mass over their lifetimes, and may thus originally have 
been even more massive. They serve to illustrate that, even in the
Milky Way, the distinction between globular and open clusters is not
always clear-cut.

  It has been recognized for about a century that the Magellanic Clouds, and 
the LMC in particular, host a number of ``blue globular clusters'' 
\citep{shap30}. %
Among the most massive of these is NGC~1866, with a mass of around 
$10^5 \, \msun$ and an age of $\approx100$ Myr \citep{fis92,van99}. An
older example is NGC~1978 with similar mass but an age of 2--3 Gyr,
clearly demonstrating that at least some such clusters can survive
for several Gyrs.  The interaction between the LMC and the Milky Way has 
probably affected the star formation history of the LMC, which is
known to be bursty with major peaks in the star formation rate correlating
with perigalactic passages \citep{smeck02}. 
One might argue, then, that the formation of YMCs in the LMC could be induced 
by interaction with the Milky Way.
However, the LMC is not the only example even in the Local Group of a
galaxy that hosts YMCs.  Another well-known example is M33, which does
not display evidence for a bursty cluster (and, presumably star-) 
formation history \citep{cs82,cs88}.  \citet{chan99,chan01} have 
identified many more star clusters in this galaxy, though not all are 
particularly massive.

  With the launch of HST it became possible to investigate more crowded
and/or distant systems in detail and attention started to shift towards more 
extreme starbursts, including a large number of merger galaxies (e.g.\ 
Whitmore, this volume). It is now clear that 
luminous, young star clusters often form in very large numbers in such 
galaxies, and this has led to suggestions that formation of ``massive''
star clusters might require special conditions such as large-scale
cloud-cloud collisitions \citep{js92}.  However, the question remains to 
be answered why some non-interacting galaxies also contain YMCs, whereas 
apparently the Milky Way does not. YMCs are now being found in an increasing 
number of non-interacting galaxies, posing a severe challenge for 
formation scenarios which require special conditions.

\section{Observations of Nearby Spirals}

  During the 1980s, some studies had already identified YMCs in a few 
galaxies beyond the Local Group \citep[][]{kc88}. 
We undertook a systematic, ground-based study of 21 nearby spirals, aiming at 
identifying cluster systems and further investigating which factors might lead 
to the formation of YMCs \citep{lr99}. 
Generally lacking sufficient resolution to identify clusters as spatially
resolved objects, our candidate lists were compiled based on $UBV$ photometry, 
selecting compact objects with $\bv<0.45$ and $M_V$ brighter than $-9.5$ (for 
$\ub<-0.4$) or $-8.5$ (for $\ub>-0.4$). We also required that the objects
had no H$\alpha$ emission.  The $\bv$ limit excluded most 
foreground stars, while the $M_V$ limit was designed to minimise the
risk that individual, 
luminous stars in the galaxies would contaminate the sample. As the 
mass-to-light ratios of star clusters are highly age dependent, the magnitude 
cut does not translate to a well-defined mass limit, but most clusters 
selected in this way have masses $\ga10^4 \, \msun$.  Our survey would 
probably pick up a few clusters in the Milky Way. In the LMC, 8 clusters
in the \citet{bica96} catalogue pass our selection criteria.

  We found a surprising variety in the numbers of YMCs in the galaxies.
Some galaxies, such as NGC~45, NGC~300 and NGC~3184 contained
hardly any clusters passing our selection criteria, but in others we found 
more than a hundred.  The two 
most cluster-rich galaxies were NGC~5236 (M83) and NGC~6946, both of which 
are also known for their very high supernova rates and surface 
brightnesses, indicative of very active star formation. Following 
\citet{har91}, we defined the \emph{specific luminosity} of young star 
clusters as
\begin{equation}
  T_L(U) = 100 \times \frac{L_{\rm clusters}}{L_{\rm galaxy}}
\end{equation}
where $L_{\rm clusters}$ and $L_{\rm galaxy}$ are the total $U$-band 
luminosities of clusters and their host galaxy. The $T_L(U)$
turned out to correlate strongly with the host galaxy area-normalised 
star formation rate \citep{lr00}, as if bound star clusters
form more efficiently in higher-SFR environments. Here, it is
important to note that our sample excludes the very youngest clusters,
which are often located in crowded regions in spiral arms where they are 
difficult to identify with ground-based imaging. Therefore, it is probably 
better to think of $T_L(U)$ as a \emph{survival-} rather than a formation 
efficiency. In fact, most stars probably form in clusters, both in normal
galaxies such as the Milky Way \citep[Carpenter this volume]{ll03} 
and in the mergers like the Antennae (Fall, this volume). The fraction of 
those clusters which remain bound may vary, however.

  While $T_L(U)$ may be a useful measure of the overall richness of a 
cluster system, it does not provide any information about possible variations 
in the cluster mass distributions, and in particular, whether some galaxies 
form a higher proportion of \emph{massive} clusters than others. This question 
still remains largely unanswered, because mass distributions are difficult to 
derive observationally.  In order to convert observed cluster
luminosities to masses, the M/L ratios need to be known.  These, in turn,
depend strongly on the cluster ages, which cannot be reliably estimated
without photometry in ultraviolet passbands. This is mostly a consequence of 
the fact that the colours of young clusters are dominated by hot stars,
where the Rayleigh-Jeans approximation applies for passbands centered
at optical wavelengths.  A ``poor man's'' solution is to look at luminosity- 
rather than mass functions, bearing in mind the first are not necessarily 
identical to the latter. 


In \citet{lar02}, archive HST data were used to analyse the luminosity 
functions (LFs)
of star clusters in 6 nearby spirals which had previously been studied from 
the ground. The LFs were generally found to be consistent with power-laws
with slopes between $-2$ and $-2.4$, somewhat steeper than the value found 
for Milky Way open clusters by \citet{van84}, though not extending nearly as
deep. For young clusters in M66, \citet{dk02} found a LF slope of $-2.53$.
%
If the luminosity function has 
a universal power-law form which is populated at random, one would predict 
a strong correlation between the total number of clusters in a galaxy and 
the luminosity of the brightest cluster.  Such a relation is indeed observed 
and has a slope, normalisation and scatter similar to those expected from 
sampling statistics arguments \citep{bhe02,lar02,whit03}. 
However, it 
should be noted that a few galaxies do stand out, having clusters that are 
much too bright for the total number of clusters in those galaxies.  Notable 
examples are NGC~1569 and NGC~1705, both of which are dominated by 
1--2 highly luminous clusters. The possibility remains open that these
clusters formed by a special mechanism, but the issue needs to be 
investigated in more detail.

\begin{figure}
\plotone{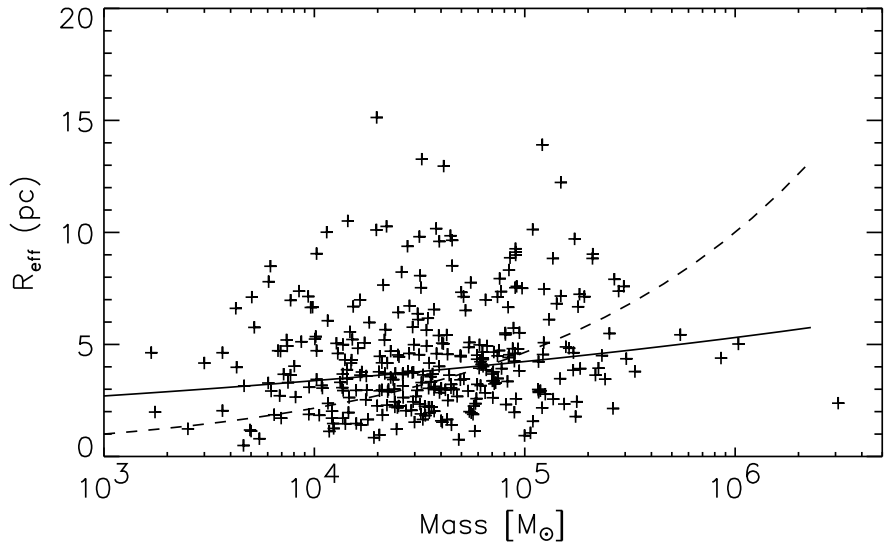}
\caption{\label{fig:reff_mass}Effective radius versus mass for
  clusters in a sample of spiral galaxies. The solid line is a
  least-squares fit to the data points while the dashed line 
  illustrates a constant-density relation ($\reff \propto M^{1/3}$).}
\end{figure}

  HST archive data for a larger sample (17) of nearby spirals were analysed
in \citet{lar04}. The main aim here was to study the structural parameters 
and investigate possible correlations with age, mass or other cluster 
properties. The clusters were modelled using `EFF' profiles of the type 
shown by \citet{eff87} to fit LMC clusters:
\begin{equation}
  P(r) = \left[1+(r/r_c)^2\right]^{-\gamma/2}
  \label{eq:eff}
\end{equation}
Structural parameters were obtained from fits to WFPC2 images and combined 
with $UBVI$ photometry from ground-based imaging.  Fig.~\ref{fig:reff_mass} 
shows the cluster half-light radii (\reff) versus masses estimated from 
Bruzual \& Charlot SSP models. Only clusters with $\gamma>2$ are included in 
this plot, as \reff\ is undefined for $\gamma\leq2$. The majority of clusters 
have half-light radii of 3--4 pc albeit with a fairly large scatter.  
Interestingly, this is similar to the effective radii of Galactic and 
extragalactic old globular clusters.  The dashed line shows the relation 
corresponding to a constant cluster density ($\reff \propto M^{1/3}$), while 
the solid line is a least-squares fit to the data. No strong correlation 
between \reff\ and mass is observed.  A formal fit to the data yields 
$\reff \propto M^{0.10\pm0.03}$, similar to the $R \propto L^{0.07}$ 
relation found for young clusters in the merger remnant NGC~3256 by 
\citet{zepf99}. The observation that no strong correlation exists between
cluster size and mass implies that high-mass clusters generally have
much higher stellar \emph{densities} than low-mass clusters, a fact that
may have important implications for theories for cluster formation.

%
%
%

\section{Dynamical mass estimates}

  If the cluster ages are known, luminosities can be converted to masses 
using simple stellar population models and assuming a stellar initial mass 
function (IMF). An alternative approach is to obtain dynamical mass 
estimates by measuring the internal velocity dispersions and cluster sizes
and applying the virial theorem.  The dynamically derived M/L ratios can 
then be compared with SSP models for different IMFs, providing a potentially 
useful method to constrain the IMF.  The line-of-sight velocity dispersion 
$v_x$, mass $M_{\rm vir}$ and projected half-light radius \reff\ are related 
as
\begin{equation}
  M_{\rm vir} \, = \, a \frac{v_x^2 \, \reff }{G}
  \label{eq:mvir}
\end{equation}
where $a\approx10$.  In practice, however, there are many caveats to this 
method, both theoretical ones (assumption of velocity isotropy, virial 
equilibrium, effects of mass segregation and binaries), and practical ones: 
For a mass of $10^5 \, \msun$ and $\reff= 3$ pc, the line-of-sight velocity 
dispersion is less than 4 km/s. 
The red supergiants which provide most of the 
lines useful for velocity dispersion measurements have macroturbulent 
velocities on the order of 10 km/s, with a scatter of perhaps 
1--2 km/s \citep{gt87}.  
Since the velocity dispersions usually have to be derived from 
integrated light, it 
is clear that this method is limited to relatively massive objects. Even
if spectra of sufficient resolution to resolve the line broadening 
($\lambda/\Delta\lambda\ga50.000$) and S/N could be obtained, an exact
match between the template stars used to derive the velocity dispersions
and those present in the cluster becomes increasingly critical as the
cluster mass decreases. 
Additionally, the clusters have to be close enough 
that reasonably reliable size estimates can be obtained, although these 
are less critical since $M_{\rm vir}$ scales only linearly with \reff . 

\begin{table}
\caption{\label{tab:cprop}Properties of 3 young stellar clusters in NGC~5236
  and NGC~6946}
\smallskip
{\footnotesize
\begin{center}
\begin{tabular}{lccccccc} \hline
Cluster     &  $D$   &     \reff    &  $M_V$  &  $A_B$  & Log(age)   &  $v_x$      &   $M_{\rm vir}$ \\
            &  Mpc   &       pc     &   mag   &   mag   &   yr       &  km/s       &   $10^5$ \msun       \\ \hline
N5236-502   &   6.0  &  $7.6\pm0.2$ & $-11.6$ &  1.0   & $8.0\pm0.1$ & $5.5\pm0.2$ & $5.2\pm0.8$ \\
N5236-805   &   4.5  &  $2.8\pm0.2$ & $-12.2$ &  1.0   & $7.1\pm0.2$ & $8.1\pm0.2$ & $4.2\pm0.7$ \\
N6946-1447  &   4.5  & $10.2\pm2.9$ & $-14.2$ &  1.3   & $7.1\pm0.1$ & $8.7\pm0.1$ & $18\pm6$ \\ \hline
\end{tabular}
\end{center}
}
\end{table}

\begin{figure}[!ht]
\plotfiddle{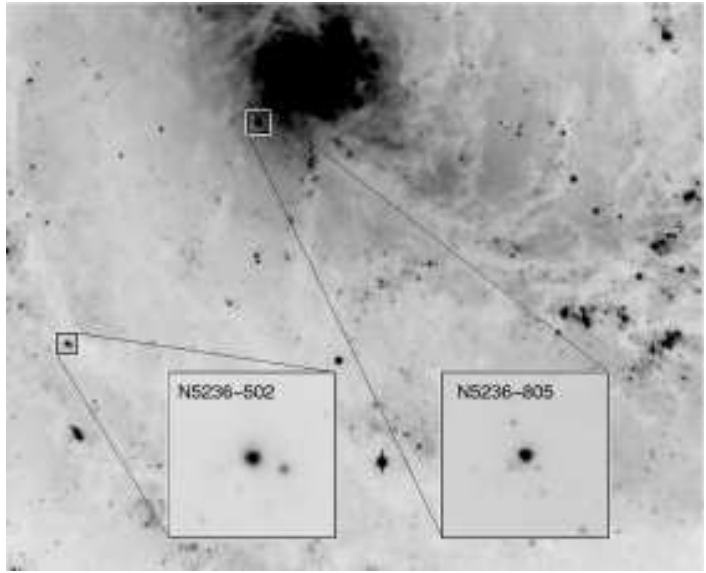}{75mm}{0}{130}{130}{-150}{0}
\caption{\label{fig:cl5236}Two massive clusters in NGC~5236 (Larsen
  \& Richtler 2004; in prep.). $V$-band
  image from the FORS2 instrument on VLT, with inserts showing
  HST/WFPC2 images of the two clusters.}
\end{figure}

\begin{figure}[!ht]
\plotfiddle{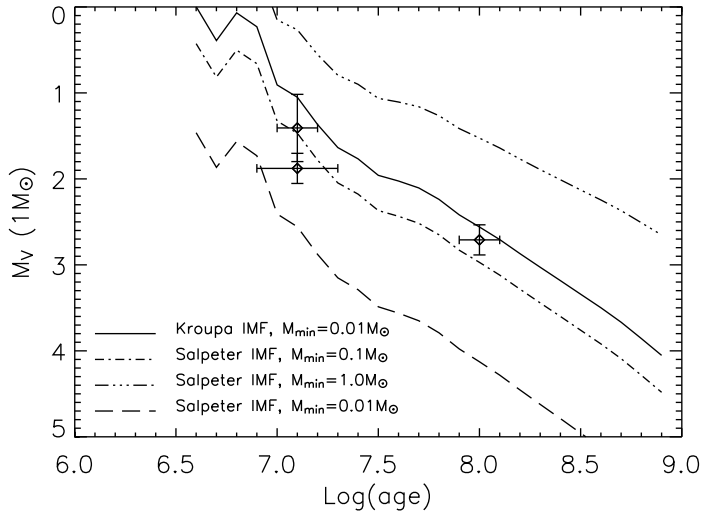}{70mm}{0}{60}{60}{-200}{-220}
\caption{\label{fig:pmtol}Comparison of observed mass-to-light ratios
  for three YMCs in NGC~6946 and NGC~5236 with model predictions
  for Salpeter and Kroupa IMFs}
\end{figure}

  Several groups have obtained dynamical mass estimates for extragalactic 
young star 
clusters, sometimes with hints of non-standard IMFs (e.g.\ \citet{sg01};
see also Mengel, this volume). In our sample of spiral galaxies, we found a 
few clusters for which dynamical mass estimates appeared feasible. Critical 
selection criteria were that the clusters be reasonably well isolated, so that 
the spectroscopic observations would not be contaminated by neighbouring 
objects, and that they have HST imaging for reliable size measurements.  
Objects which satisfy these criteria include one cluster in NGC~6946 which was 
observed with the HIRES spectrograph on the Keck I telescope 
\citep{lar01}, and two clusters in NGC~5236 (Fig.~\ref{fig:cl5236}), observed 
with UVES on the ESO 
VLT \citep{lr04}.  Structural parameters are available for all clusters from 
HST imaging, and ages and reddenings were estimated by comparing ground-based 
$UBVI$ colours with Bruzual \& Charlot SSP models.  

  Basic properties for the three clusters are summarised in 
Table~\ref{tab:cprop}.  They all have masses greater than $10^5\msun$,
well in excess of those of the most massive young LMC clusters.  Even if our
reference frame had been the LMC rather than the Milky Way, we would still 
have characterised these clusters as ``massive''.  In Fig.~\ref{fig:pmtol}, 
the observed $V$-band M/L ratios are compared with SSP model predictions for 
various IMFs. These SSPs were computed by populating stellar isochrones from 
the Padua group \citep{gir00} according to the IMFs indicated in the
figure legend, i.e.\ a \citet{kroupa02} IMF and \citet{salp55}-type
IMFs with lower mass cut-offs at 0.01 \msun, 0.1 \msun\ and 1.0 \msun .
The Kroupa IMF nominally extends down to 0.01 \msun, but the lower
cut-off is of little importance as the slope below 0.08 \msun\ is
very shallow.  
 Within the error bars, the three clusters all appear consistent with a 
``standard'' Kroupa-like IMF. In particular, there is no evidence for
an excess of high-mass stars in any of these clusters.  We checked the 
curves in Fig.~\ref{fig:pmtol} against the Bruzual \& Charlot models, which
are available for Salpeter IMF truncated at 0.1 $\msun$ and found
very similar results, with differences at the level of 0.1 mag at most.
It should be emphasized that this method does not constrain 
the exact shape of the IMF. Power-law IMFs with a shallow slope, for example, 
would mimick the effect of a Salpeter IMF with a high-mass cut-off.  


\section{Concluding remarks}

  It is becoming increasingly clear that ``massive'' star clusters can form 
in a wide variety of galaxies, and not just in mergers or otherwise disturbed 
galaxies. With the possible exception of some dwarf galaxies, the luminosity 
distributions of young star clusters generally appear to be power-laws. 
If cluster luminosities are sampled at random from a power-law 
distribution, the most luminous clusters will naturally be rare, but so far 
there is no evidence for a statistically significant upper cut-off. In other 
words, very luminous (and massive) clusters appear to form whenever clusters 
form in large 
numbers.  
This is illustrated by the fact that young star clusters with masses up to 
$\sim10^6 \, \msun$ have been identified in the disks of several apparently 
normal, isolated spiral galaxies with rich cluster systems.  These galaxies, 
such as NGC~5236, NGC~6946 are characterised by high star formation rates, 
but these do not generally appear to be triggered by interactions with 
other galaxies. Dynamical mass estimates are now available for a small
number of these clusters, and the mass-to-light ratios are compatible
with standard Kroupa-type IMFs. There is every reason to be optimistic 
that important clues to the formation of classical globular clusters may be
obtained by studying their younger counterparts in the Local Universe.

\acknowledgements

  I am grateful to my colleagues and collaborators who have contributed
valuable help and insight, especially T. Richtler and J. Brodie.

\end{document}